\documentclass[prc,aps,
showpacs,
nofootinbib,floatfix,
twocolumn,
preprintnumbers]{revtex4}
\usepackage{latexsym}

\usepackage{amsmath, mathrsfs}
\usepackage[latin1]{inputenc}
\usepackage[dvips]{graphicx}

\newcommand{\qstil}{\widetilde{Q}_\mathrm{s}}
\newcommand{\rp}{R_p}
\newcommand{\qscoord}{{Q_\mathrm{s,coord.}}}
\newcommand{\qscoordsqr}{{Q_\mathrm{s,coord.}^2}}

\newcommand{\xt}{\mathbf{x}_T}
\newcommand{\yt}{\mathbf{y}_T}

\newcommand{\kt}{{\mathbf{k}_T}}

\newcommand{\nabt}{\boldsymbol{\nabla}_T}

\newcommand{\ktil}{\tilde{k}}

\newcommand{\ud}{\, \mathrm{d}}

\newcommand{\tr}{\, \mathrm{Tr} \, }
\newcommand{\nc}{{N_\mathrm{c}}}

\newcommand{\half}{\frac{1}{2}}

\newcommand{\cf}{C_\mathrm{F}}
\newcommand{\ca}{C_\mathrm{A}}
\newcommand{\df}{d_\mathrm{F}}
\newcommand{\da}{d_\mathrm{A}}
\newcommand{\nr}[1]{(\ref{#1})} 
\newcommand{\ra}{R_A}

\newcommand{\gev}{\ \textrm{GeV}}
\newcommand{\fm}{\ \textrm{fm}}

\newcommand{\qs}{{Q_\mathrm{s}}}
\newcommand{\lqcd}{\Lambda_{\mathrm{QCD}}}
\newcommand{\as}{\alpha_{\mathrm{s}}}

\newcommand{\fig}{Fig.~}
\newcommand{\eq}{Eq.~}

\begin{document}

\author{T. Lappi}

\email{tuomas.lappi@cea.fr}

\affiliation{
Institut de Physique Th\'eorique, B\^at. 774, CEA/DSM/Saclay, 91191
 Gif-sur-Yvette Cedex, France
}

\title{Wilson line correlator in the MV model: 
relating the glasma to deep inelastic scattering}

\preprint{arXiv:0711.3039 [hep-ph]}
\preprint{SPhT-T07/146}

\begin{abstract}
In the color glass condensate framework the saturation scale measured in deep 
inelastic scattering of high energy hadrons and nuclei
can be determined from the correlator of Wilson lines in the hadron wavefunction.
These same Wilson lines give the initial condition of the classical field 
computation of the initial gluon multiplicity and energy density in a heavy ion 
collision.
In this paper the Wilson line correlator in both adjoint and fundamental 
representations is computed using exactly the same numerical procedure
that has been used to calculate gluon production in a heavy ion collision.
In particular the discretization of the longitudinal coordinate has a large 
numerical effect on  the relation between the color charge density parameter 
$g^2\mu$ and the saturation scale $\qs$. Our result for this relation is
$\qs \approx 0.6 g^2 \mu$, which results in the 
classical Yang-Mills value for the ``gluon liberation
coefficient'' $c\approx 1.1$.
\end{abstract}

\maketitle

\section{Introduction}
\label{sec:intro}

A useful description of the hadron or nucleus wavefunction at high energy
is to view the small $x$ degrees of freedom as classical color fields radiated
by classical static color sources formed by the large $x$ degrees of freedom
\cite{McLerran:1994ni,McLerran:1994ka,McLerran:1994vd}.
This description, known as the color glass condensate (for reviews see e.g.
\cite{Iancu:2003xm,Weigert:2005us}), provides
a common framework for understanding both small $x$ deep inelastic scattering (DIS)
and the initial stages of relativistic heavy ion collisions, both of which 
can be understood in terms of Wilson lines of the classical color field.
The cross section for small $x$ DIS can be expressed in terms of the correlator
of two Wilson lines in the fundamental representation (i.e. the dipole cross section),
and the initial condition
for the classical fields that dominate the first fraction of a fermi of a 
heavy ion collision is determined by these same Wilson lines.
The inverse of the correlation length of these Wilson lines is known as the 
\emph{saturation scale} $\qs$.
The  dipole cross section, can be determined from the 
dipole model fits to DIS data on protons 
\cite{Golec-Biernat:1998js,Golec-Biernat:1999qd,Stasto:2000er,Iancu:2003ge,%
Kowalski:2003hm}
and nuclei \cite{Freund:2002ux,Armesto:2004ud} or extensions
from the proton to the nucleus using a parametrization of the nuclear geometry 
 \cite{Levin:2001et,Gotsman:2002yy,Levin:2002fj,Kowalski:2003hm,Kowalski:2006hc,%
Kowalski:2007rw}. 
On the other hand there is a large
body of both analytical 
\cite{Kovner:1995ts,Kovner:1995ja,Gyulassy:1997vt,Dumitru:2001ux,%
Kovchegov:1997ke,Fukushima:2007ja,Fries:2006pv} 
and numerical classical Yang-Mills (CYM) \cite{% 
Krasnitz:1998ns,Krasnitz:1999wc,Krasnitz:2000gz,Krasnitz:2001qu,%
Krasnitz:2003jw,%
Krasnitz:2002mn,Lappi:2003bi,Lappi:2004sf,Lappi:2006hq} computations of 
the ``Glasma'' \cite{Lappi:2006fp} fields in the initial stages of relativistic
heavy ion collisions.

The aim of this paper is to relate the parameters of these two types applications
of color glass condensate ideas of the to each other more precisely. This is done by
computing the Wilson line correlator in the 
McLerran-Venugopalan (MV) model 
\cite{McLerran:1994ni,McLerran:1994ka,McLerran:1994vd}
using exactly the same numerical method that has been used to compute the 
initial transverse energy and multiplicity in a heavy ion collision. By doing this
we can relate saturation scale $\qs$, whose numerical value can be determined from 
fits to DIS data, to the color charge density $g^2 \mu$ that determines the initial conditions for a heavy ion collision. The calculation relating these two parameters
has been done analytically by several authors \cite{Jalilian-Marian:1997xn,% 
Kovchegov:1996ty,Kovchegov:1998bi,McLerran:1998nk,Gelis:2001da,Blaizot:2004wu,% 
Blaizot:2004wv}. The procedure used to construct the MV model Wilson lines
in this paper is the same as used in the numerical computations of the Glasma fields
and differs from these analytical computations in two ways. 
Firstly, as noted also in Ref.~\cite{Fukushima:2007ki}, the analytical 
computation is done by spreading out the color source in rapidity, while in the 
numerical computations this has not been done. We shall see that this introduces a 
factor of 2 difference in the actual numerical relation between $\qs$ and 
$g^2\mu$. 
Secondly the analytical result for the relation between $\qs$ and $g^2 \mu$
depends logarithmically on an infrared cutoff that must be used in an intermediate
stage of the computation, whereas in most of the numerical work the only such 
cutoff has been the size of the system.
We shall also discuss the uncertainty arising from the non-Gaussian functional 
form of the Wilson line correlator and argue that it introduces 
an additional ambiguity at the 10\% level.
While the uncertainty from these 
aspects is parametrically unimportant (a constant or a logarithm), they must still 
be better understood in order to increase the predictive power of the calculations.

The logic of this paper is that, instead of treating the color charge density 
$g^2\mu$ in the ``Glasma'' calculations as a free phenomenological parameter, one
should be able to relate it exactly, even the constant under the logarithm, to the
saturation scale $\qs$ measured in DIS experiments. When DIS measurements
are used to determine the value of the saturation scale, choosing what treatment
of the rapidity direction to use in the MV model is mostly a matter of convenience 
as long as the value of $g^2\mu$ used is consistent with this chosen implementation.
Let us note that 
our concern here is not as much the effects of high energy evolution on the wavefunction,
but the parametrization of the region $x \sim 0.01$ relevant 
for central rapidities at RHIC, which would be a reasonable initial condition
for solving the BK or JIMWLK equation. The glasma field configurations obtained
are the boost invariant fields that serve as the background for studying things
like instabilities in the classical \cite{Romatschke:2005pm,Romatschke:2006nk,%
Fukushima:2006ax} field and higher order contributions to particle production
\cite{Gelis:2006yv,Gelis:2006cr,Gelis:2007kn}.

We shall first introduce our notation for the Wilson line correlators in 
Sec.~\ref{sec:wlines}. Then our numerical results are presented in
Sec.~\ref{sec:numerics} and their implications for the interpretation of some of the
earlier phenomenological work discussed in Sec.~\ref{sec:disc}.

\section{Wilson lines and glasma fields}\label{sec:wlines}

Consider a high energy nucleus or a hadron moving along the $x^+$-axis. 
Its fast degrees of freedom can be considered as a classical color current
\begin{equation} \label{eq:current}
J^+ = g\rho(\xt,x^-),
\end{equation}
which acts as a source to a classical color field representing the slower partons
\begin{equation} \label{eq:cym}
[D_\mu,F^{\mu \nu}] = J^\nu.
\end{equation}
In the MV model the color charge density is taken to be a stochastic random variable
with a Gaussian distribution.

In covariant gauge \eq\nr{eq:cym} can be solved as
\begin{equation} \label{eq:Aplus}
A^+(x^-,\xt) = -\frac{g\rho(\xt,x^-)}{\nabt^2}.
\end{equation}
The path ordered exponential of this field gives the Wilson line in the fundamental 
representation
\begin{equation}\label{eq:wline}
U(\xt) = P e^{i \int  \ud x^- A^+}.
\end{equation}
It is this quantity that will concern us in the following.

The cross section for a virtual photon scattering off a high energy
hadron or nucleus can be expressed in terms of the dipole cross section, which
is determined by the correlator of two Wilson lines in the fundamental 
representation \cite{Rummukainen:2003ns,Weigert:2005us} 
\begin{equation} \label{eq:wlinecorr}
\widetilde{C}(\xt-\yt) = \langle  \tr U^\dag(\xt) U(\yt)\rangle,
\end{equation}
with the expectation value $\langle \rangle$ evaluated with the distribution 
of the sources.

\begin{figure}[tb]
\begin{center}
\includegraphics[width=0.4\textwidth]{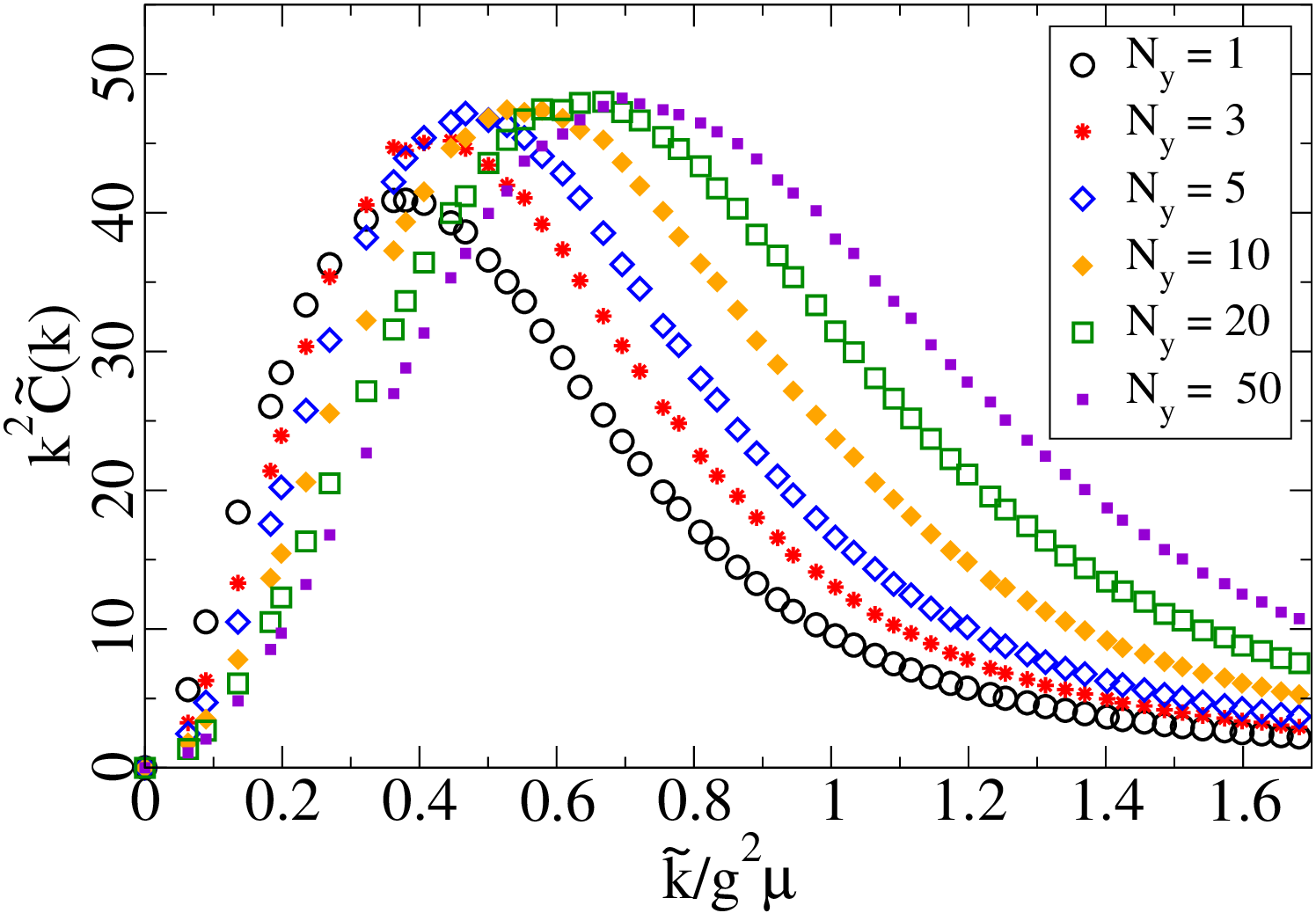}
\includegraphics[width=0.4\textwidth]{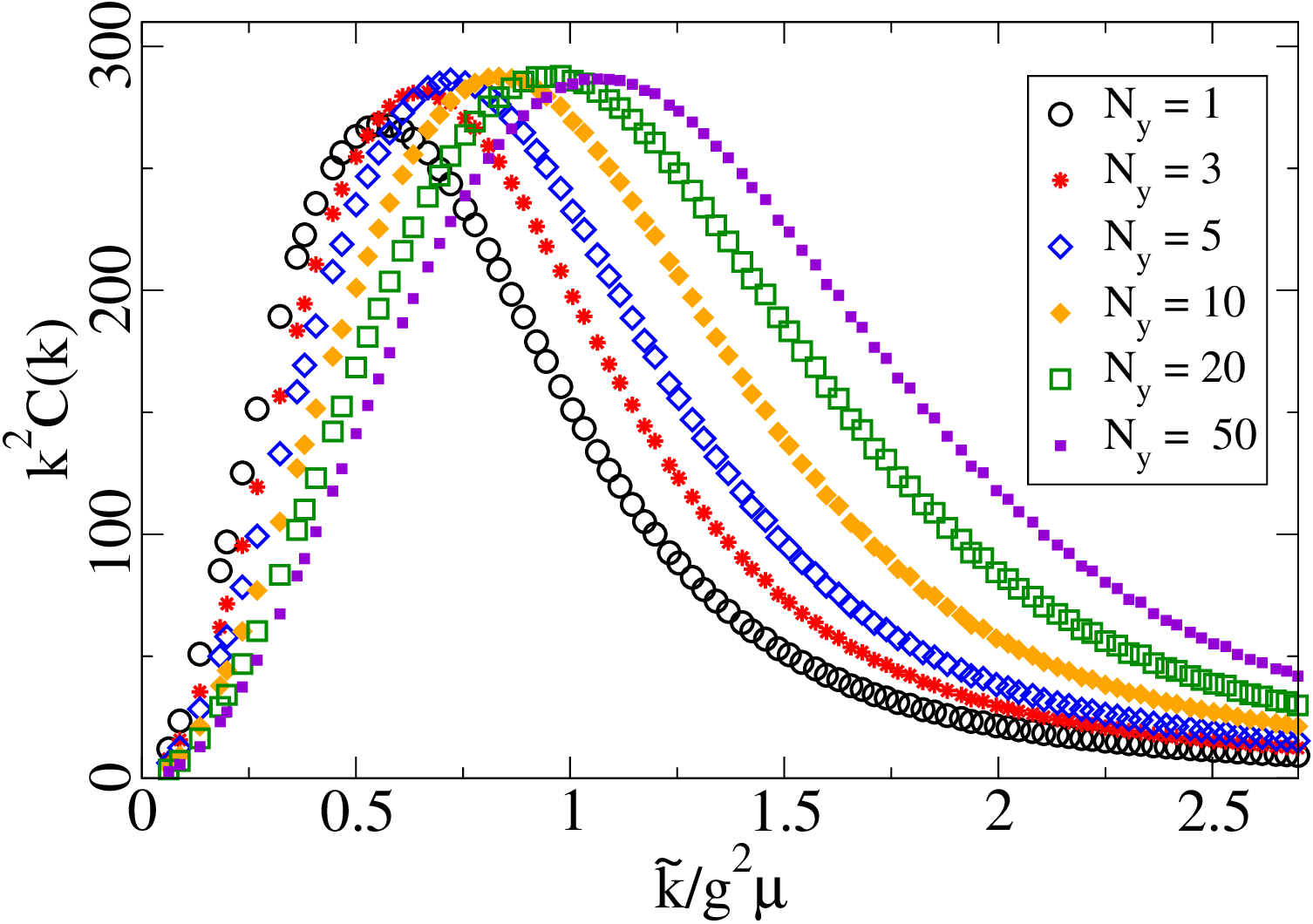}
\end{center}
\caption{
The Wilson line correlator for $g^2\mu L = 100$
and different values of $N_y$. Above: fundamental representation
$\ktil^2 \widetilde{C}(\ktil)$, below: adjoint representation
$\ktil^2 C(\ktil)$.
}\label{fig:fundadjcorr}
\end{figure}

The Wilson line in the adjoint representation is given by
\begin{equation} \label{eq:adjU}
U_{ab}(\xt) = 2\tr \left[t^a U^\dag(\xt) t^b U(\xt) \right].
\end{equation}
The correlator of adjoint representation Wilson lines
\begin{equation} \label{eq:adjwlinecorr}
C(\xt-\yt) = \langle U_{ab}(\xt) U_{ab}(\yt) \rangle
\end{equation}
is related to the gluon distribution of a nucleus 
\cite{Jalilian-Marian:1997xn,Kovchegov:1998bi,Kovchegov:2000hz}
(See Refs.~\cite{Kharzeev:2003wz,Blaizot:2004wu,Gelis:2006tb} for a discussion
on the intricacies of defining a gluon distribution in this case.)
With some algebra this the adjoint representation correlator can be related
to a higher correlator of fundamental representation Wilson lines
\begin{equation} \label{eq:adjwlinecorr2}
C(\xt-\yt) =  \left\langle  \left| \tr \left[U^\dag(\xt)U(\yt) \right] 
 \right|^2 
- 1 \right\rangle,
\end{equation}
which is the form we shall use to evaluate it numerically.

The initial conditions for the glasma fields are determined by the 
pure gauge fields (in light cone gauge) of the two colliding nuclei
\cite{Kovner:1995ts,Gyulassy:1997vt}. In terms of 
the Wilson line \nr{eq:wline} the pure gauge field of one nucleus is
\begin{equation}
A^i(\xt) = \frac{i}{g}U(\xt)\partial_i U^\dag(\xt),
\end{equation}
and the initial conditions for the glasma fields are given by the sum and
commutator of the pure gauge fields of the two nuclei.
In the numerical computation of the glasma fields 
\cite{Krasnitz:1998ns,Krasnitz:1999wc,Krasnitz:2000gz,Krasnitz:2002mn,%
Krasnitz:2001qu,Lappi:2003bi,Krasnitz:2003jw%
} 
there has been no longitudinal structure in the source, and the Wilson lines
have been constructed simply as
\begin{equation}\label{eq:numwline}
U(\xt) = \exp\left\{ -i \frac{g \rho(\xt)}{\nabt^2} \right\},
\end{equation}
with the transverse charge densities depending on a single parameter $\mu$, 
independent of $x^-$:
\begin{equation} \label{eq:deltamv}
\langle \rho^a(\xt)\rho^b(\yt)\rangle = \delta^{ab}\delta^2(\xt-\yt) g^2 \mu^2.
\end{equation}

The analytical calculation \cite{Jalilian-Marian:1997xn,% 
Kovchegov:1996ty,Kovchegov:1998bi,McLerran:1998nk,Gelis:2001da,Blaizot:2004wu,% 
Blaizot:2004wv}  of the Wilson line correlator requires that, unlike the numerical
procedure in \cite{Krasnitz:1998ns,Krasnitz:1999wc,Krasnitz:2000gz,Krasnitz:2002mn,%
Krasnitz:2001qu,Lappi:2003bi,Krasnitz:2003jw}, 
the source be extended in the $x^-$ direction
\begin{multline} \label{eq:extmvcorr}
\langle \rho^a(\xt,x^-)\rho^b(\yt,y^-)\rangle 
=
\\
 g^2 \delta^{ab}\delta^2(\xt-\yt)
\delta(x^- - y^-) \mu^2(x^-).
\end{multline}
With this longitudinal structure the Wilson line correlators can be computed
analytically up to a logarithmic
infrared cutoff that must be introduced in solving the Poisson equation~\nr{eq:Aplus}.
The result is 
\begin{eqnarray} \label{eq:extcorrs}
\widetilde{C}(\xt) &\approx& \df e^{\frac{\cf}{8 \pi} \chi \xt^2 \ln(m |\xt|)} \\
\nonumber
C(\xt) & \approx & \da e^{\frac{\ca}{8 \pi} \chi \xt^2 \ln(m |\xt|)},
\end{eqnarray}
with
\begin{equation} \label{eq:extmv}
\chi = g^4 \int \ud x^- \mu^2(x^-).
\end{equation}
The dimensions and Casimirs of the two representations 
in \eq\nr{eq:extcorrs}
are $\da = \nc^2 - 1$, $\df = \nc$,
$\ca = \nc$ and $\cf = (\nc^2-1)/2\nc$.
It could be argued that the cutoff $m$ should be $\sim \lqcd$. In any case, running 
coupling and confinement effects are not included in this treatment and the cutoff
cannot be consistently defined within this calculation. When looking at length scales
$|\xt| \ll 1 / m$ results depend very weakly on this cutoff; in the lattice 
calculation it can be replaced by the finite size of the lattice.
It would be very tempting to identify $\mu^2$, the source strength of the delta function
source,  appearing
in \eq\nr{eq:deltamv}, with the integral over the spread distribution 
$\mu^2(x^-)$ of
\eq\nr{eq:extmv}, but as we will see in the following, this correspondence is 
not exact.\footnote{It is relatively easy to see that the
identification of $\int \ud x^- \mu^2(\xt,x^-)$ with $\mu^2(\xt)$ of 
\eq\nr{eq:deltamv}
 would be exact in the Abelian
case or in the large $\nc$ limit in which the terms resulting from the noncommutative
nature of $\rho$ are suppressed.}

Note that the form \nr{eq:extcorrs} is compatible with the expectation that in
the large $\nc$ limit the four point function in \eq\nr{eq:adjwlinecorr2}
factorizes into a product of two point functions and
\begin{equation} \label{eq:largeNc}
\lim_{\nc \to \infty} C(\xt) = \widetilde{C}^2(\xt).
\end{equation}

\begin{figure}
\begin{center}
\includegraphics[width=0.45\textwidth]{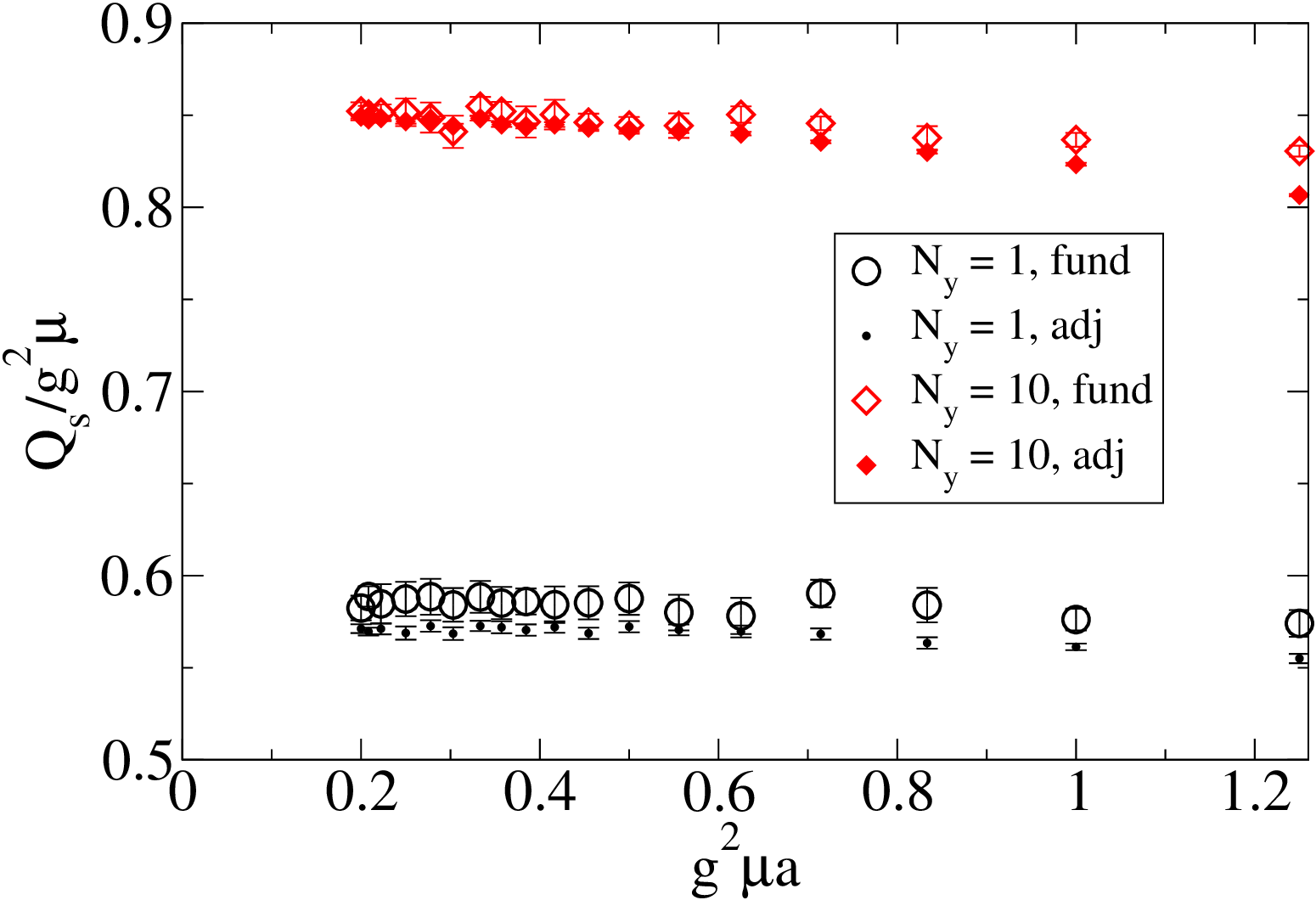}
\end{center}
\caption{The lattice spacing dependence of the saturation scales  $\qs$ and $\sqrt{\frac{\ca}{\cf}}\qstil$
for $g^2\mu L = 100$ and different $N_y$. The continuum limit is the $g^2\mu a = 0$
axis on the left.
}\label{fig:contlim}
\end{figure}

\begin{figure}
\begin{center}
\includegraphics[width=0.45\textwidth]{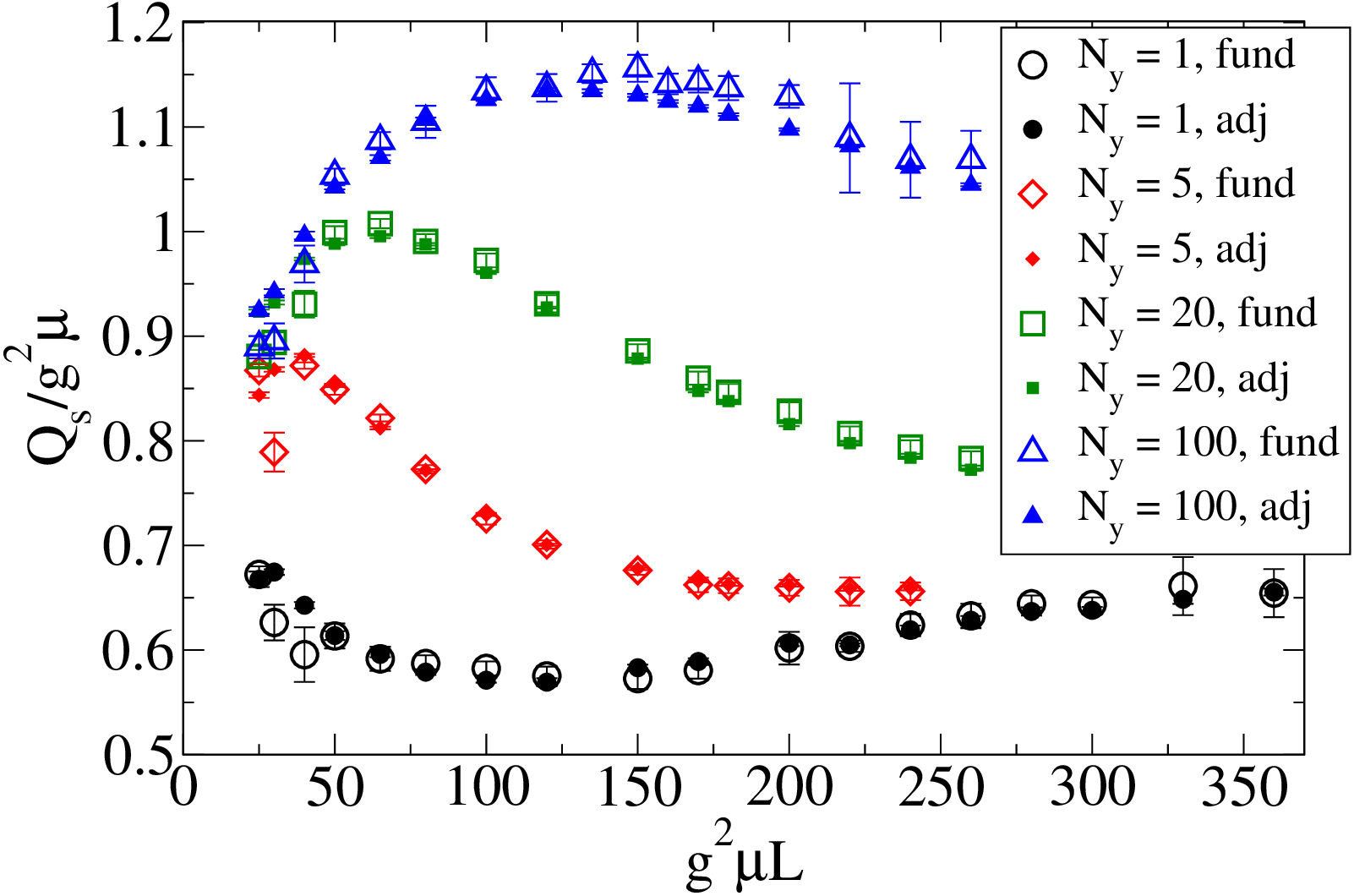}
\end{center}
\caption{
The dependence of the adjoint and fundamental representation
saturation scales $\qs$ and $\sqrt{\frac{\ca}{\cf}}\qstil$
on $g^2\mu L$ for $g^2\mu a = 0.5$ and different $N_y$.
}\label{fig:mul}
\end{figure}

\section{Numerical procedure and results} \label{sec:numerics}

The Wilson lines used in the numerical calculation of the Glasma fields \cite{%
Krasnitz:1998ns,Krasnitz:1999wc,Krasnitz:2000gz,Krasnitz:2001qu,Krasnitz:2003jw,%
Krasnitz:2002mn,Lappi:2003bi,Lappi:2004sf}
are SU(3) matrices defined on the sites of a 2 dimensional discrete lattice 
corresponding to the transverse plane. As in most of these 
calculations, we shall consider a square lattice with periodic boundary conditions
and an average color charge density $g^2 \mu^2$ that is constant throughout the plane. 
The Wilson lines are constructed as follows:
On each lattice site $\xt$ one constructs random color charges  with a 
local Gaussian distribution
\begin{equation}\label{eq:discrsrc}
\left\langle \rho^a_k(\xt) \rho^b_l(\yt) \right\rangle =
 \delta^{ab}  \delta^{kl}  \delta^2(\xt-\yt)
\frac{g^2 \mu^2}{N_y},
\end{equation}
with the indices $k,l=1,\dots,N_y$ representing a discretized longitudinal coordinate.
The numerical calculations so far have been done using $N_y=1$, whereas the derivation
of the analytical expression of the correlator, \eq\nr{eq:extcorrs} are derived with an 
extended source, corresponding to the limit $N_y \to \infty$. Our normalization is chosen 
so that
\begin{equation}
\sum_{k,l} \left\langle \rho^a_k(\xt) \rho^b_l(\yt) \right\rangle =
 \delta^{ab}    \delta^2(\xt-\yt)
g^2 \mu^2.
\end{equation}
The Wilson lines are then constructed from the sources \nr{eq:discrsrc} by solving a 
Poisson equation and exponentiating:
\begin{equation}\label{eq:uprod}
U(\xt)  = \prod_{k=1}^{N_y} \exp\left\{ -i g \frac{\rho_k(\xt)}{\nabt^2 + m^2}\right\}.
\end{equation}
Here we have introduced an infrared regulator $m$ for inverting the Laplace operator.
This is the same regulator as the one appearing in the analytical expression
\eq\nr{eq:extcorrs}. 
For large $N_y$ the charge densities $\rho_k$in \eq\nr{eq:discrsrc} become
small, and the individual elements in the product \nr{eq:uprod} approach identity.
This is precisely the procedure that leads  in the $N_y \to  \infty$ limit to
the continuum  path ordered exponential \nr{eq:wline}.

\begin{figure}
\begin{center}
\includegraphics[width=0.45\textwidth]{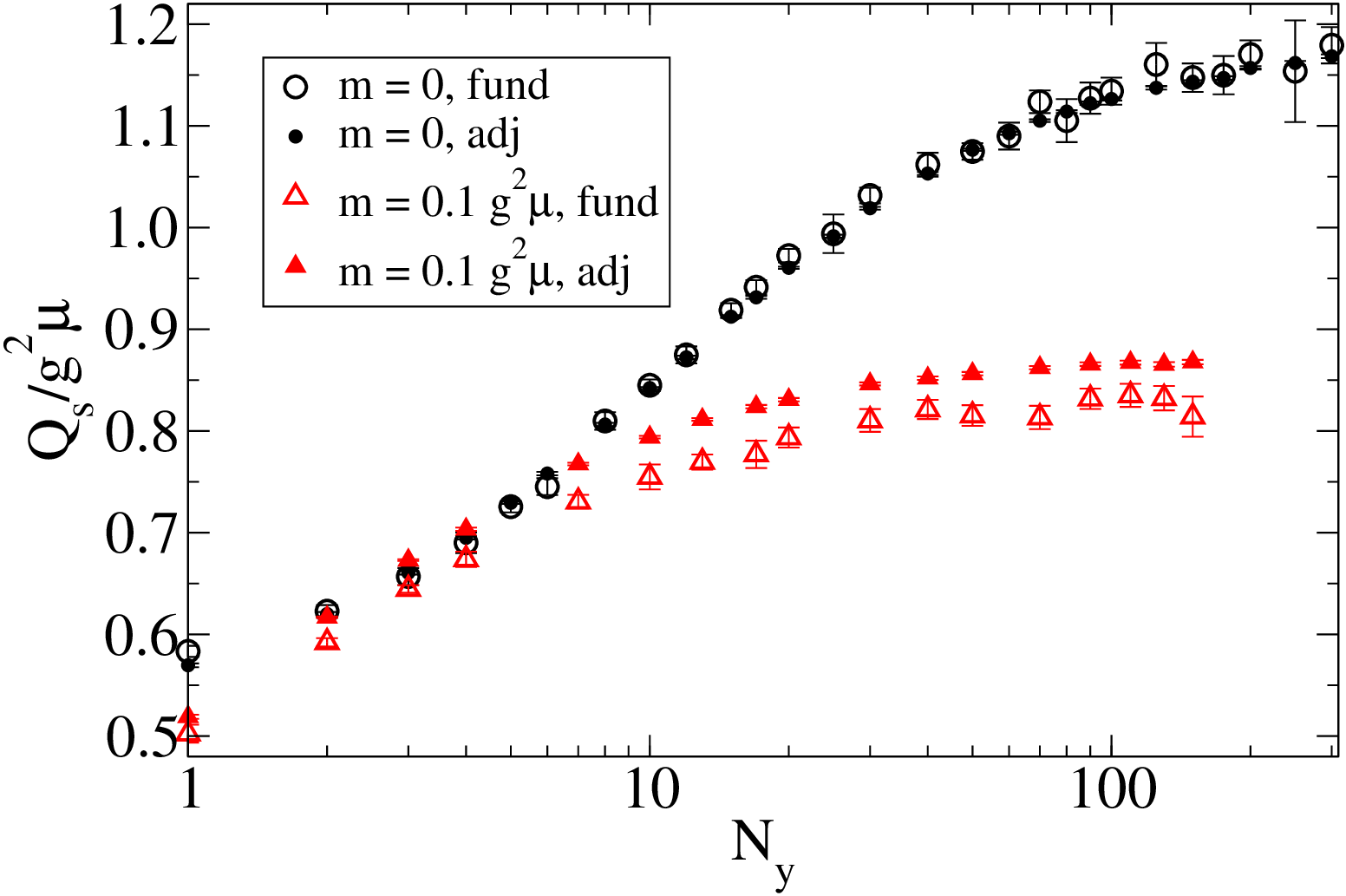}
\end{center}
\caption{
Dependence on $N_y$ of the saturation scales  $\qs$ and $\sqrt{\frac{\ca}{\cf}}\qstil$ 
for $g^2\mu L = 100$
and $g^2\mu a = 0.5$, shown for $m=0$ and $m=0.1 g^2 \mu$.
}\label{fig:ny}
\end{figure}

To summarize, our calculation depends on the following parameters: 
\begin{itemize}
\item $g^2 \mu$, determining the color charge density.
\item $N_y$, the number of points in the discretization of the longitudinal ($x^-$ 
or rapidity) direction. 
\item The infrared regulator $m$.
When $m=0$, as in most of the results presented, the Poisson equation
is solved by leaving out the zero transverse momentum mode. This procedure 
corresponds to an infrared cutoff given by the size of the system.
\item The lattice spacing $a$.
\item The number of transverse lattice sites $N_\perp$, giving the size of the lattice
$L = N_\perp a$.
\end{itemize}
Of the parameters $a$, $g^2\mu$ and $m$, only the dimensionless combinations
$g^2\mu a$ and $m a$ appear in the numerical calculation, and 
the continuum limit $a\to 0$ is taken by letting $N_\perp \to \infty$ so that
$g^2 \mu a \to 0$ and $g^2 \mu L = g^2 \mu a N_\perp$ remains constant.
What we are looking at is relatively infrared quantity and thus should converge very well 
in the continuum limit. Based on the analytical calculation we may expect a logarithmic 
dependence of the saturation scale on $g^2\mu/m$ or, for $m=0$, on $g^2\mu L$.

By Fourier transforming the Wilson lines we can than construct the momentum space
correlators in the adjoint and fundamental representations, $C(\kt)$ and 
$\widetilde{C}(\kt)$ respectively. These correlators, averaged over the polar angle, for different values of $N_y$
are plotted in Fig.~\ref{fig:fundadjcorr} as a function of 
\begin{equation}
\ktil = \frac{2}{a} \sqrt{\sum_{i=1}^2 \sin^2 \left( k_i a / 2\right)}.
\end{equation}
For small momenta the correlators look like Gaussians, which is the form 
used in the ``GBW'' fit of DIS data in 
Refs.~\cite{Golec-Biernat:1998js,Golec-Biernat:1999qd,Stasto:2000er}.
For large momenta there is a power law tail $1/\kt^4$ that differs from the
original GBW fits, but resembles more closely the form required to match smoothly
to DGLAP evolution for large $Q^2$ \cite{Bartels:2002cj}.

\begin{figure}
\begin{center}
\includegraphics[width=0.4\textwidth]{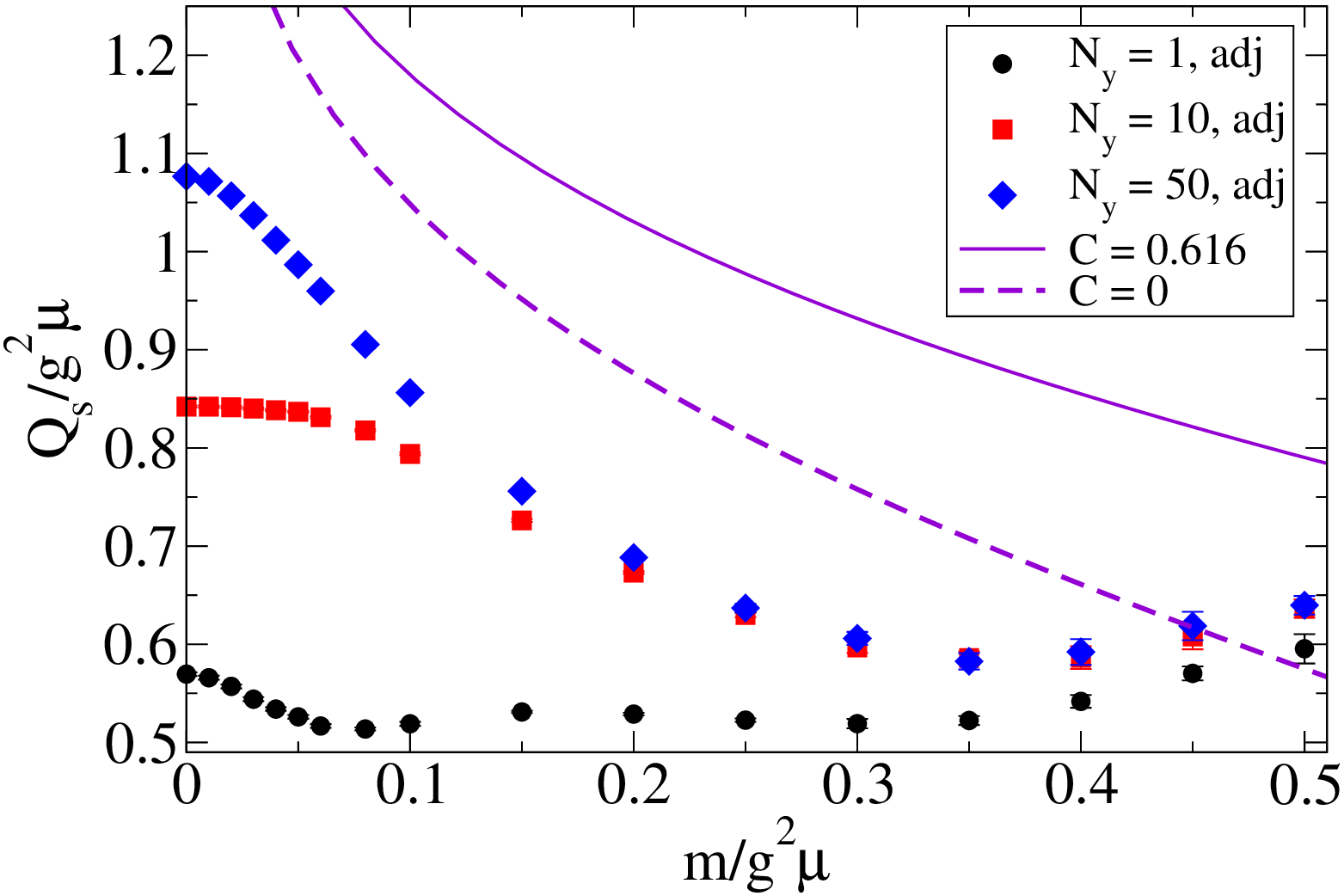}
\end{center}
\caption{
Dependence on the regulator $m$ of the saturation scale
$\qs$  for 
$g^2\mu L = 100$. The solid line is the expected logarithmic 
dependence, \eq\nr{eq:analqscst} with the constant $C = 0.616$
and the dashed one with $C=0$.
}\label{fig:mass}
\end{figure}

We define the numerically measured saturation scales as follows.
The scale $\qs$ is determined by the adjoint representation 
Wilson line correlator as the momentum $\ktil_\mathrm{max}$ corresponding
to the maximum of $\ktil^2 C \left(\kt\right)$. This normalization 
in terms of the adjoint representation corresponds to that
of Refs.~\cite{Kovchegov:1998bi,Kovchegov:2000hz}. 
Similarly, from the maximum of
the fundamental representation correlator $\ktil^2 \widetilde{C}\left(\kt\right)$ we 
define the fundamental representation
saturation scale $\qstil$ as $\qstil^2 = \ktil_\mathrm{max}^2$.
Our definition of the saturation scale 
 is not sensitive to the exact shape of the correlator for very large
or small transverse momenta, and for a Gaussian correlator it reproduces
the GBW saturation scale as $1/R^2_0 = \qstil^2.$
The saturation scale is expected to scale according to the Casimir
of the representation, meaning $\qstil^2 \approx \frac{\cf}{\ca}\qs^2$.
In the plots (Figs.~\ref{fig:contlim}, \ref{fig:mul}, \ref{fig:ny}
and \ref{fig:mass}) we shall rescale $\qstil$ by this color factor 
to make the validity of this scaling clearer.

We first check the lattice spacing dependence of our result. Figure~\ref{fig:contlim}
shows that, as expected, the ratio $\qs/g^2\mu$ depends in fact so 
little on the lattice spacing that we will in the following not perform any continuum 
extrapolation for this quantity. The dependence of $\qs$ on the lattice size
through the combination $g^2\mu L$ (without 
the additional infrared cutoff $m$) is shown
in Fig.~\ref{fig:mul}. The values used in the numerical
computations of the glasma fields \cite{% 
Krasnitz:1998ns,Krasnitz:1999wc,Krasnitz:2000gz,Krasnitz:2001qu,Krasnitz:2003jw,% 
Krasnitz:2002mn,Lappi:2003bi,Lappi:2004sf,Lappi:2006hq} 
correspond to $N_y=1$ and 
$g^2\mu L \sim 100$ in Fig.~\ref{fig:mul}, with $\qs \approx 0.57 g^2\mu$.

\begin{figure}
\begin{center}
\includegraphics[width=0.4\textwidth]{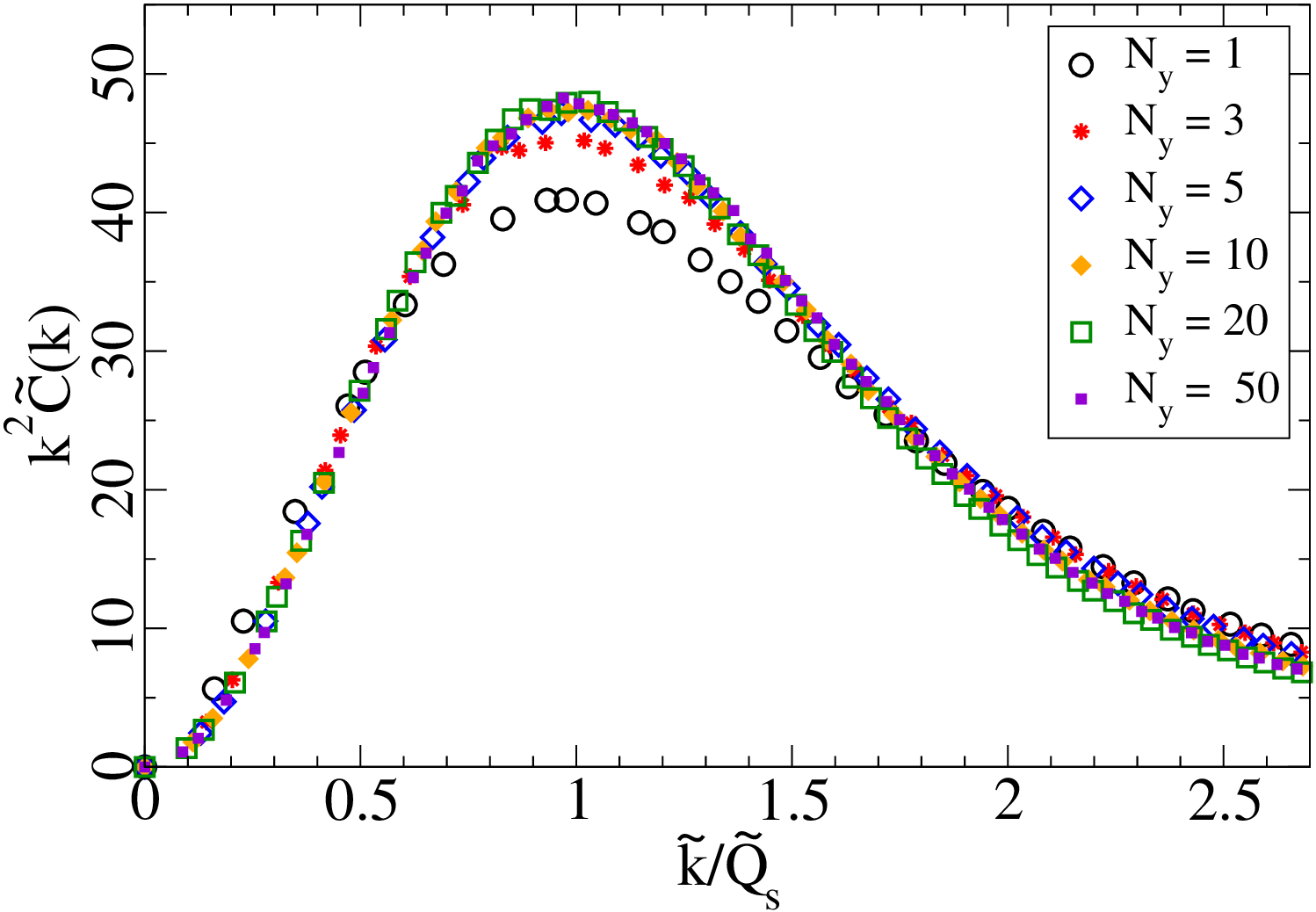}
\includegraphics[width=0.4\textwidth]{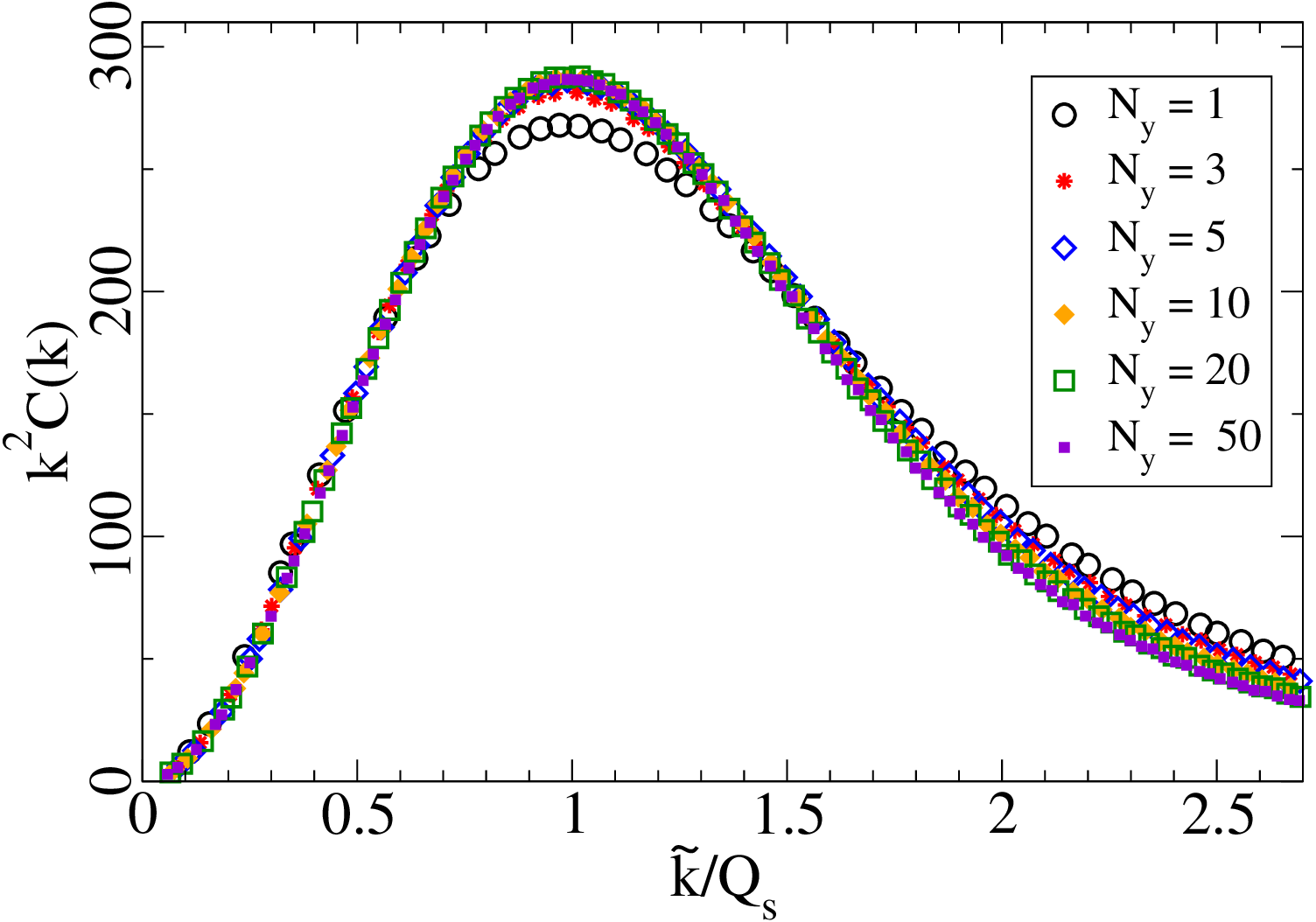}
\end{center}
\caption{
The same  fundamental representation 
Wilson line correlator as in \fig\ref{fig:fundadjcorr}
 plotted as a function of the scaling variable
$\ktil/\qs$. 
Above: fundamental representation $\ktil^2 \widetilde{C}(\ktil)$  vs.
$\ktil/\qstil$, below: adjoint representation 
$\ktil^2 C(\ktil)$ vs. $\ktil/\qs$.
}\label{fig:uspectscqs}
\end{figure}

Figure~\ref{fig:ny} shows the dependence of $g^2\mu/\qs$ on the number of points
used to discretize the longitudinal direction, $N_y$. 
When $m=0$, i.e. the infrared singularity is regulated only by leaving out the
zero mode, there is approximately a factor of 
two difference between $\qs = 0.57 g^2\mu$ for $N_y=1$
(the numerical CYM prescription)
and $\qs \approx 1.15 g^2 \mu$ for $N_y \to \infty$ 
(the analytical computation of the dipole cross section)\footnote{Because the initial
energy density $\epsilon$ of the glasma is proportional to $\qs^4/g^2$, this
factor of 2 could be an explanation of
 the factor of 16 difference in $\epsilon/(g^2\mu)^4$ observed in
Ref.~\cite{Fukushima:2007ki}.}.
When a regulator $m=0.1 g^2\mu$ is introduced the dependence on $N_y$ 
is weaker,  which can also be seen in Fig.~\ref{fig:mass}.
In Fig.~\ref{fig:uspectscqs}  we show the same correlators
as in Fig.~\ref{fig:fundadjcorr} as a function of $\ktil/\qs$
instead of $\ktil/g^2\mu$. One sees that the correlator has a scaling form 
independent of $N_y$; from which only the $N_y=1$ result deviates slightly.
This suggests that, as argued in Sec.~\ref{sec:intro}, once the appropriate
relation between $\qs$ and $g^2\mu$ is used, the physical results depend very little on
$N_y$. Thus no significant change to the numerical CYM results should be expected
if the calculations were repeated using a different treatment of the longitudinal 
coordinate in the source $\rho$.

\begin{figure}
\begin{center}
\includegraphics[width=0.4\textwidth]{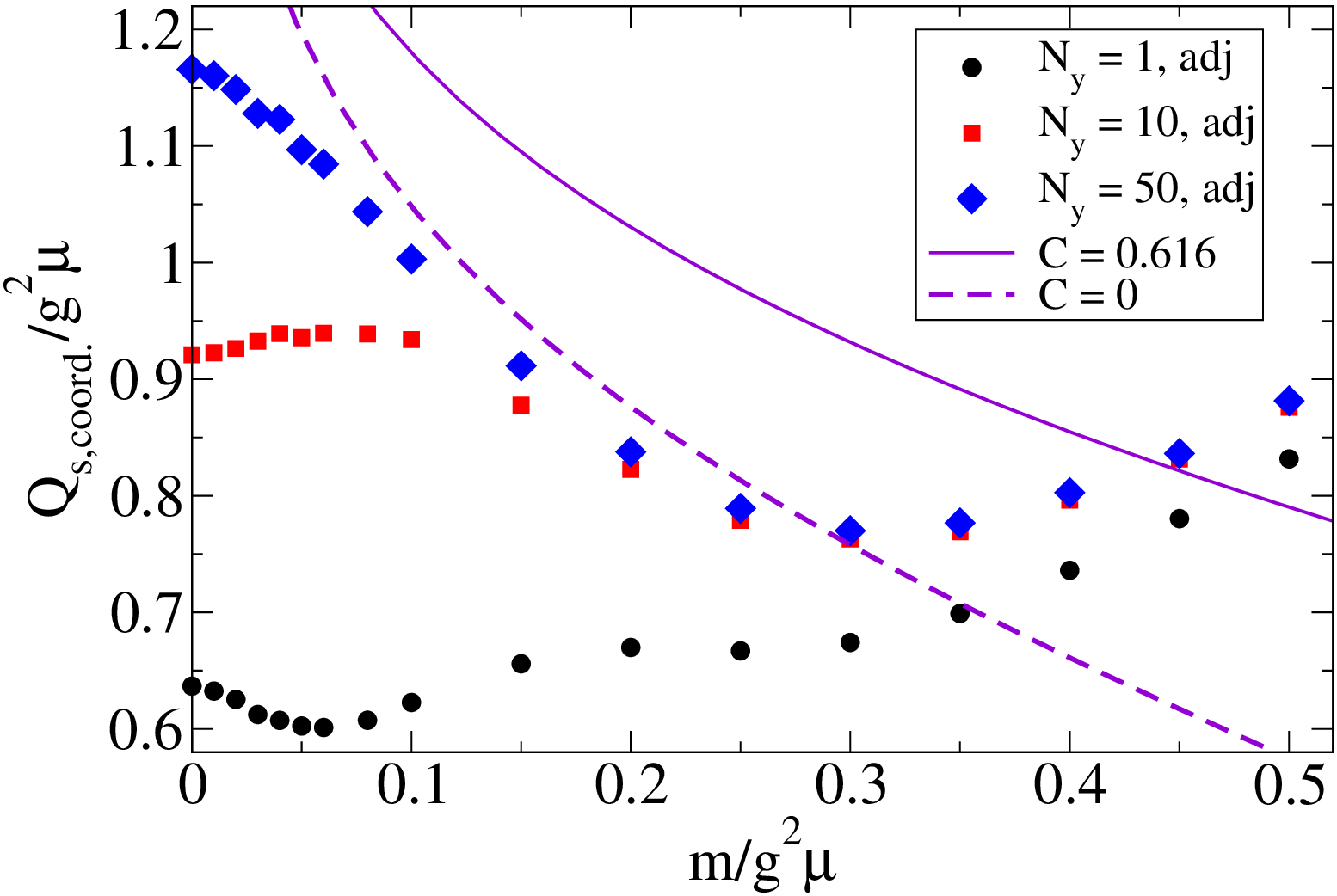}
\end{center}
\caption{
Dependence on the regulator $m$ of the coordinate space 
saturation scale $\qscoord$  for 
$g^2\mu L = 100$. The solid line is the expected logarithmic 
dependence, \eq\nr{eq:analqscst} with the constant $C = 0.616$
and the dashed one with $C=0$.
}\label{fig:masscsp}
\end{figure}

Explicitly regulating the infrared behavior with a mass scale $m$ makes it
possible to compare the numerical result to the analytical one of \eq\nr{eq:extcorrs}.
If one introduces an infrared scale $m$ as in \eq\nr{eq:numwline} and replaces
$\ln \left( m \left|\xt \right| \right)$ with $-\ln \left( g^2 \mu / m \right)$ 
in the coordinate space correlator it becomes a Gaussian.  
Fourier transforming this one obtains the estimate 
\begin{equation}\label{eq:analqs}
\frac{\qs^2 }{\left( g^2 \mu \right)^2} 
\approx 
\frac{\ca}{\cf} \frac{\qstil^2}{ \left( g^2 \mu \right)^2} 
\approx 
\frac{\ca}{2 \pi } \left[ \ln \frac{g^2 \mu}{m} + \half + \ln 2 
-\gamma_\mathrm{E} \right].
\end{equation}
Because of the replacement $|\xt| \sim 1/g^2 \mu$ there is still an uncertainty
in the constant term.  In \fig\ref{fig:mass} we plot the numerical result for 
$\qs/g^2\mu$ as a function of $m/g^2\mu$ compared to the estimate
\begin{equation}\label{eq:analqscst}
\frac{\qs^2 }{\left( g^2 \mu \right)^2} 
= 
\frac{\ca}{2 \pi } \left[ \ln \frac{g^2 \mu}{m} + C \right],
\end{equation}
with values $C = \half + \ln 2 -\gamma_\mathrm{E} \approx 0.616$ and $C=0$.
In an intermediate range of $m/g^2\mu$ and for a large enough value of $N_y$
(recall that the analytical result corresponds to $N_y\to \infty$) the 
behavior of $\qs/g^2\mu$  is similar, but the normalization different.

\begin{figure}
\begin{center}
\includegraphics[width=0.45\textwidth]{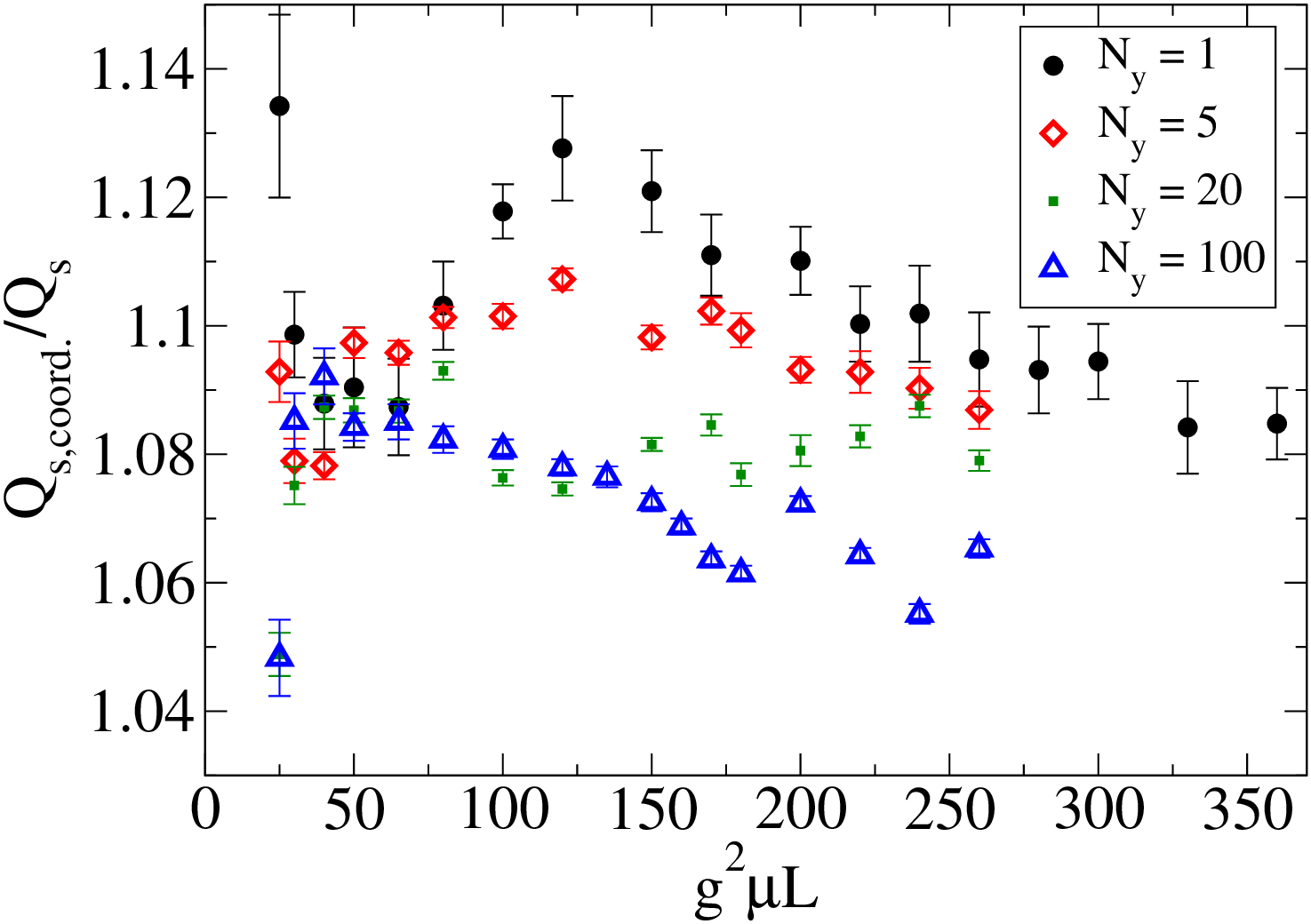}
\end{center}
\caption{
The ratio of the two definitions of the saturation scale, 
the coordinate space definition ($\qscoord$) and
the momentum space definition used in most of this paper ($\qs$).
}\label{fig:gsqrmulrat}
\end{figure}

Another common way to define the saturation scale is in terms of the 
coordinate space correlator $C(\xt)$, because this is the object appearing 
in the calculation of most observables in DIS. Kowalski and 
Teaney~\cite{Kowalski:2003hm} define the saturation scale
$\qscoord$ from the condition that $C(\xt) = \da e^{-1/2}$ 
at $\xt^2 = 2/\qscoordsqr$. Note that the definition in 
Ref.~\cite{Kowalski:2007rw} where the same IPsat model 
is used differs slightly:
 $C(\xt) = \da e^{-1/4}$  at $\xt^2 = 1/\qscoordsqr$. 
 This definition 
can also be used in the numerical CYM computation, most straightforwardly 
 by Hankel-transforming the correlator $C(|\kt|)$ back into coordinate space.
As shown in \fig\ref{fig:masscsp}, using this definition is closer to 
the analytical estimate \eq\nr{eq:analqscst}. The ratio of 
$\qscoord$ to our original definition of $\qs$ for different values
of $g^2\mu L$ and $N_y$ is plotted in \fig\ref{fig:gsqrmulrat}. The difference
between the two definitions is of the order of 10\% with small variations. 
One must emphasize here that for an exactly Gaussian Wilson line correlator 
(the GBW form) the two definitions would be equal. They differ in the MV model, 
because the correlator is not Gaussian. Thus if one tries to 
determine $g^2\mu$ from a comparison to the experimental DIS data
using GBW-type fits, which is one of the alternatives we consider in the 
next section, the
ambiguity in the definition of $\qs$ leads to a 10\% uncertainty in the value
of $g^2\mu$.

\section{Discussion} \label{sec:disc}

Let us finally use the results of the previous section for $\qs/g^2\mu$ and studies of 
DIS data to estimate the relevant value of $g^2\mu$ for RHIC physics.
In deep inelastic scattering the variables $x$ and $Q^2$ are precisely defined, and
the saturation scale is a function of $x$, typically $\qs^2 \sim x^{-\lambda}$ with
$\lambda \approx 0.3.$ In the 
context of a heavy ion collision one is in fact, at a fixed energy and rapidity, 
summing up gluons 
produced at different transverse momenta and thus related to partons of different $x$
in the nuclear wavefunction. The value of $x$ at which to evaluate the saturation scale
must therefore be some kind of effective $x_\textrm{eff}$, depending on the typical
transverse momentum of 
the produced gluons, 
\begin{equation}
x_\textrm{eff} \sim \frac{\langle p_T \rangle}{\sqrt{s}} \sim \frac{\qs}{\sqrt{s}}.
\end{equation}
This introduces an additional uncertainty
into our attempt to determine the color charge density based on the deep inelastic 
scattering data; by varying $\half \qs/\sqrt{s} < x_\textrm{eff} < 2 \qs/\sqrt{s}$ 
we get an uncertainty of the order of 5\%.
 
A simple starting point for our estimate 
is the GBW fit \cite{Golec-Biernat:1998js,Golec-Biernat:1999qd},
where the  proton saturation scale (in the fundamental representation, which is convenient
for DIS) is parametrized as
\begin{equation}
\qstil^2 = \frac{1}{R_{0 \ \mathrm{GBW}}^2} 
= Q_0^2 (x/x_0)^{\lambda}.
\end{equation}
The result of the fit
including charm quarks gives  $\lambda = 0.277$ and $x_0 = 0.41 \cdot 10^{-4}$, 
with the one redundant parameter chosen as $Q_0 = 1 \gev$, while the fit without
charm makes $\qs$ approximately 30\% larger.

This result must then be extended to
finite nuclei. Let us denote the nuclear modification
of $\qs$ by $g(A) \equiv \qs_A^2/\qs_p^2$. The most straightforward theoretical
expectation for the nuclear dependence would be $g(A) = A^{1/3}$.
Freund et al.~\cite{Freund:2002ux} perform a fit of the form $g(A) = A^\delta$ 
to the available nuclear DIS data and obtain $\delta=1/4$. 
Taking into account modifications
to the $A^{1/3}$ behavior of the nuclear radius leads Armesto et al.~\cite{Armesto:2004ud} 
to consider a fit of the form
$g(A) \sim A \rp^2/\ra^2 =
C \left[A/(A^{1/3}-0.77 A^{-1/3})^2 \right]^{3\delta}$ with the result
$\delta \approx 0.42$ and $C \approx 0.5$. 
Although for asymptotically large nuclei this would imply
$g(A) \sim A^{0.42}$, for the physical case $A \lesssim 200$ the nuclear modification
factor $g(A)$ obtained in Ref.~\cite{Armesto:2004ud} is actually less than $A^{1/4}$.

\begin{table}
\begin{tabular}{|l||r|r|}
\hline
 & \multicolumn{1}{|l|}{$\qs^\textrm{RHIC}$} & \multicolumn{1}{|l|}{$g^2 \mu$} \\
 \hline \hline
Naive $A^{1/3}$		& $1.7\gev$ & $3.0\gev$ \\
\hline
 $A^{1/4}$ \cite{Freund:2002ux}	& $1.4\gev$ & $2.5 \gev$ \\
\hline
$C A^{4/9}$  \cite{Armesto:2004ud}	& $1.3\gev$ & $2.2 \gev$ \\
\hline
 IPsat, $\sim C A^{1/3} \ln A$ \cite{Kowalski:2007rw} & $1.1\gev$ & $2.0 \gev$ \\
\hline
\end{tabular}
\caption{Results for the adjoint representation saturation scale
from extrapolations of DIS data to RHIC central  rapidity 
kinematics.} \label{tab:qsrhic}
\end{table}

A more detailed description of the saturation scale in a nucleus can be obtained by the
IPsat model~\cite{Kowalski:2003hm,Kowalski:2006hc,Kowalski:2007rw}.
HERA data and the DGLAP equations are used to parametrize the dipole cross section for 
a proton. Taking into account the fluctuations in the 
positions of the nucleons in the nucleus within a realistic nuclear geometry
leads to a nuclear dipole cross section, from which also the saturation scale
can be determined. As shown in Ref.~\cite{Kowalski:2007rw} this picture
leads to a good parameter free description of all the existing small $x$ eA data.
The result is a more realistic picture of
an impact parameter dependent saturation scale also influenced by DGLAP evolution,
where  the nuclear geometry leads to a $g(A)$ that can roughly 
be understood as a $C A^{1/3}$-like dependence (with $C < 1$) enhanced by a 
logarithmic increase in $A$ resulting from the DGLAP evolution.
Because scattering off nuclei is less dominated by the dilute edge than in the proton,
the typical $\qs$ (conveniently taken as corresponding to 
$b_\textrm{med.}$,  the median impact parameter in deep inelastic scattering) is closer
to the maximal $\qs$ in the nucleus than in the proton. We shall use here for 
gold the value at $b_\textrm{med.} \approx 4.2 \fm $ and at
at $x=0.005 \approx \qs/\sqrt{s}$.

Table \ref{tab:qsrhic} summarizes 
the estimated saturation scales for calculating the 
classical field at central rapidity in RHIC based on the different
fits explained above. 
The table also shows the corresponding values of $g^2\mu = \qs/0.57$.

To set this result in perspective let us briefly recall the result of the numerical CYM 
computations. The energy and multiplicity per unit rapidity 
can be parametrized as
\begin{eqnarray}
\frac{\ud N}{\ud \eta} &=& \frac{(g^2\mu)^2 \pi \ra^2}{g^2}f_N 
\label{eq:fn}
\\
\frac{\ud E_T}{\ud \eta} &=& \frac{(g^2\mu)^3 \pi \ra^2}{g^2}f_E,
\end{eqnarray}
where the numerical result is $f_E\approx 0.25$ and $f_N\approx 0.3$
(see in particular Refs.~\cite{Lappi:2003bi,Krasnitz:2003jw} for the result).
This leads to the estimate that, for central rapidity at RHIC,
$1.3\gev \lesssim g^2\mu \lesssim 2.1 \gev$.
The lower limit comes from the requirement that the energy per unit rapidity 
has to be at least as large initially as it is finally. 
In boost invariant hydrodynamical (``Bjorken'') flow energy is transferred
from central rapidities to the fragmentation region by $p \ud V$ work thus decreasing the energy at central rapidities, decreasing $\ud E_T/\ud y$.
It is very hard to imagine a process that would increase the energy 
at midrapidity and thus the measured final transverse energy 
gives a lower limit to the initial energy and to $g^2\mu$.
The upper limit follows from the requirement that the number of gluons
in the initial state should be less or equal to that of hadrons in the 
final state. In ideal hydrodynamical flow the two are related by entropy 
conservation, and nonequilibrium processes should increase entropy and consequently the multiplicity during the evolution, not decrease it. 
The measured hadron multiplicity thus gives an upper limit on the initial multiplicity and $g^2\mu$.
Quark pair production \cite{Gelis:2004jp,Gelis:2005pb} or in general
higher order
processes would generically increase the initial multiplicity
for a given $g^2\mu$ and thus decrease the upper limit for $g^2 \mu$ 
below $2\gev$.
The only overlap region between these estimates and the DIS based ones in 
Table~\ref{tab:qsrhic} is around $g^2\mu \approx 2 \gev$. This is also   
close to the estimate, based on large $x$ parton distribution functions,
of Ref.~\cite{Gyulassy:1997vt}
that was  used in the CYM calculations 
of Refs.~\cite{Krasnitz:1999wc,Krasnitz:2000gz,Krasnitz:2001qu}.
On the DIS side the value $g^2\mu \approx 2 \gev$ agrees very well 
with the estimate using the IPsat model (see table~\ref{tab:qsrhic}), but our
present understanding of gluon production seems to be in contradiction 
with the other, more naive, DIS fits. On the 
ion-ion collision side, to a large initial gluon multiplicity and rapid
equilibration, leaving little room for 
higher order effects or additional gluon production during thermalization.

One feature of the numerical CYM calculations has been the apparent 
small fraction of the gluons in the initial nuclear wavefunction that are
``freed'' in the collision \cite{Kovchegov:2000hz,Mueller:2002kw}. 
The liberation coefficient $c$, 
introduced by A. Mueller~\cite{Mueller:1999fp},
is defined by writing the produced gluon multiplicity as
\begin{equation} \label{eq:libc}
\frac{\ud N}{\ud^2 \xt \ud y} = c \frac{\cf \qs^2}{2 \pi^2 \as}.
\end{equation}
With \eq\nr{eq:fn} this leads to
\begin{equation} \label{eq:libc2}
c = \frac{\pi f_N}{2 \cf} \left( \frac{g^2\mu}{\qs} \right)^2.
\end{equation}
The original expectation was that $c$ should be of order unity. The
analytical calculation by Y. Kovchegov~\cite{Kovchegov:2000hz}
gave the estimate  $c \approx 2\ln2 \approx 1.4$. Using the  
formula \nr{eq:analqs} for the ratio $\qs/g^2\mu$ led to the interpretation
\cite{Baier:2002bt,Mueller:2002kw} that the
CYM result would be $c \approx 0.5$.
We now see that when $\qs/ g^2 \mu$ is computed consistently with the numerical 
calculation the resulting CYM value for the
liberation coefficient is $c \approx 1.1$. We must emphasize that, because 
$c$ is defined in terms of the physical multiplicity  and the physical correlation
length $\qs$, there is no large logarithmic or $N_y$ uncertainty in the
result $c\approx 1.1$. The non-Gaussianity of the MV model correlator, as
seen in the differing coordinate and momentum space results for $\qs$,
does introduce an ambiguity at the 10\% level.

Let us summarize the major sources of error in estimating the relevant value of the 
saturation scale for RHIC physics from the DIS data. We have already
mentioned the questions of the Wilson line correlator not being exactly of the GBW
form, the exact value of $x$ to use and the considerable variance in the
estimates of $A$ dependence of $\qs$.  It is also possible that including
a more realistic description of the transverse coordinate dependence of the 
saturation scale~\cite{Gotsman:2002yy,Kowalski:2003hm,Krasnitz:2002mn,%
Lappi:2006xc,Kowalski:2007rw}
 in the CYM calculation will have an impact on the 
gluon multiplicity and energy in an ion-ion collision, modifying our previous discussion.
The solution to the problems related to the shape of the correlator can be solved 
by using the actual solution of the BK or JIMWLK equations to understand
both DIS data (as is done in Ref.~\cite{Iancu:2003ge}) and to calculate
the Glasma fields. Confirming the calculations 
like that of Ref.~\cite{Kowalski:2007rw} relating 
the saturation descriptions of the proton and a
nucleus will require more experimental input in the form of more data
on small $x$ DIS on nuclei.
Finally and perhaps most importantly, the influence of quantum corrections and
instabilities of small rapidity-dependent fluctuations is
not yet understood quantitatively.

\acknowledgments{
The author would like to thank R. Venugopalan for numerous discussions and 
comments on the manuscript.}

\bibliographystyle{h-physrev4mod}
\bibliography{spires}

\end{document}